\documentclass{mearimproc}

\usepackage{graphicx}
\usepackage[super]{natbib}

\newcommand{\ltsim}{\mbox{{\raisebox{-0.4ex}{$\stackrel{<}{{\scriptstyle\sim}}
$}}}}
\title{Multi-wavelength extragalactic surveys and the role of MeerKAT and SALT}
\author{Matt J.Jarvis$^{1,2}$}{$^1$Centre for Astrophysics Research, Science \& Technology Research Institute, University of Hertfordshire, Hatfield, Herts, AL10 9AB, UK }{$^2$Physics Department, University of the Western Cape, Cape Town, 7535, South Africa \\M.J.Jarvis@herts.ac.uk}

\begin{document}

\maketitle
{In these proceedings I discuss a range of surveys that are currently underway at optical, near-infrared and far-infrared wavelengths that have large components accessible to both the Southern African Large Telescope (SALT) and the Meer Karoo Array Telescope (MeerKAT). Particular attention is paid to the surveys currently underway with ESO's VISTA telescope, which will provide the ideal data from which to select targets for SALT spectroscopy whilst also providing the necessary depth and photometric redshift accuracy to trace the $\mu$Jy radio population, found through the proposed MeerKAT surveys. Such surveys will lead to an accurate picture of evolution of star-formation and accretion activity traced at radio wavelengths. Furthermore, SALT spectroscopy could play a crucial role in following up {\sl Herschel} surveys with its large collecting area and blue sensitivity which occupies a niche in instrumentation on 8- and 10-m class telescopes. }


\section{Introduction}
Over the coming decade we will see the completion of a large number of major astronomical surveys covering all wavelengths across large swathes of the sky. 
Such surveys are already underway or being planned, and many more will
undoubtedly be initiated in the coming years. Here I
focus on the near-infrared surveys currently being carried out by the
ESO-Visible and Infrared Survey Telescope for Astronomy (VISTA) at Paranal, Chile, in particular the second
deepest tier of the extragalactic public surveys to be undertaken on
this telescope, namely the VISTA Deep Extragalactic Observations
(VIDEO) survey. I also briefly discuss the various other multi-wavelength
surveys that will be carried out over these regions of sky.

I also discuss two of the surveys underway with the {\sl Herschel Space Observatory}, {\sl Herschel}-ATLAS and HerMES, which when combined with shorter wavelength data will allow us to trace the obscured activity in the Universe.

Finally, I discuss how the South African Large Telescope and the MeerKAT can play a major role in advancing our understanding of the extragalactic Universe when combined with the imaging surveys underway elsewhere.

\section{Extragalactic near-infrared surveys with VISTA}\label{sec:vistasurveys}

VISTA will conduct several public surveys at near-infrared wavelengths ($0.8 - 2.4\mu$m). There are currently six approved surveys, I briefly discuss three of the surveys most relevant for extragalactic studies with the SKA and its precursors, and then discuss the VIDEO survey in greater detail, as this is the most relevant for the approved MeerKAT continuum and H{\sc i} surveys.

\noindent
$\bullet$ Ultra-VISTA is the deepest tier of the extragalactic surveys. It will survey the COSMOS field with three separate strategies. The ultra-deep survey covers 0.73~deg$^{2}$ to AB magnitudes of $Y=26.7$, $J=26.6$, $H=26.1$ and $K_{s}=25.6$, with the aim of detecting galaxies within the epoch of reionisation at $z>6$. The narrow-band survey is expected to find $\sim 30$ Ly$\alpha$ emitters at $z \sim 8.8$ to a depth of NB$_{\rm AB}=24.1$ and the wide survey will complete the coverage of the full 1.5~deg$^{2}$ COSMOS field, to depths of  $Y=25.7$, $J=25.5$, $H=25.1$ and $K_{s}=24.5$ (all AB).

\smallskip

\noindent
$\bullet$ The VISTA Kilo-degree Infrared Galaxy (VIKING) survey aims to survey two stripes at high galactic latitude in five near-infrared filters. The areas have been chosen to overlap with the regions of sky covered by the 2df galaxy redshift survey\cite{Colless01} and the optical Kilo-Degree Survey (KIDS) to be conducted with the VLT Survey Telescope, thus providing both spectroscopic redshifts up to $z\sim 0.3$ and photometric redshifts up to and above $z\sim 1$. The science aims are heavily based on this photometric redshift accuracy for cosmological studies to constrain the dark energy component in the Universe, gain more accurate mass density measurement from weak lensing, find $z>7$ quasars and trace galaxy evolution from $z\sim 1$ to the present day. These regions are also the subject of the {\sl Herschel}-ATLAS\cite{Eales10}.
It is apparent that any $\sim 1000$~deg$^{2}$ survey from the SKA-precursors should target these regions to maximise the scientific productivity of both continuum and H{\sc i} surveys.
\smallskip

\noindent
$\bullet$ The VISTA Hemisphere Survey (VHS) aims to survey the rest of the southern sky which is not being covered by the other VISTA surveys. Although much shallower than the other surveys, it will provide data covering $\sim 18000$~deg$^{2}$ to a depth of at least $K_{\rm s} \sim 19.8$ and $J\sim 20.9$.
The majority of this area will also be covered by various surveys conducted at optical wavelengths, namely the Dark Energy Survey (DES; https://www.darkenergysurvey.org/) and the VST-ATLAS survey, in addition to the various galactic plane studies. Such a survey has wide-ranging scientific goals and its complementarity to any large area SKA precursor survey is obvious. If the SKA precursors progress as planned toward the full SKA then it is plausible that H{\sc i} redshifts could be obtained for a significant fraction of the VHS galaxies, allowing detailed investigations between the H{\sc i} and the stellar properties of the galaxies.
\smallskip


\section{VISTA Deep Extragalactic Observations (VIDEO) survey}\label{sec:VIDEO}

The  VIDEO survey aims to obtain both deep and wide near-infrared
observations over well-studied contiguous fields covering
$3-4.5$~deg$^{2}$ each. The fields chosen for this are derived from
the equatorial and southern fields of the {\sl Spitzer} Wide-area Infrared
Extragalactic Survey (SWIRE\cite{Lonsdale03}), namely $\sim
3$~deg$^{2}$ within the Elais-S1 region, $\sim 4.5$~deg$^{2}$ within
the XMM-Newton Large Scale Structure (XMM-LSS) survey and 4.5~deg$^{2}$
around the Chanda Deep Field South (CDFS).  An approved companion survey, the {\sl Spitzer} Extragalactic Representative Volume survey (SERVs), has also been allocated 1400~hours with the {\sl Spitzer-}warm mission. These two surveys will enable galaxy evolution to be traced from the epoch of reionization through to the present day and as a function of environmental density. 

\begin{figure*}[ht!]
\centering
\includegraphics[width=0.8\linewidth]{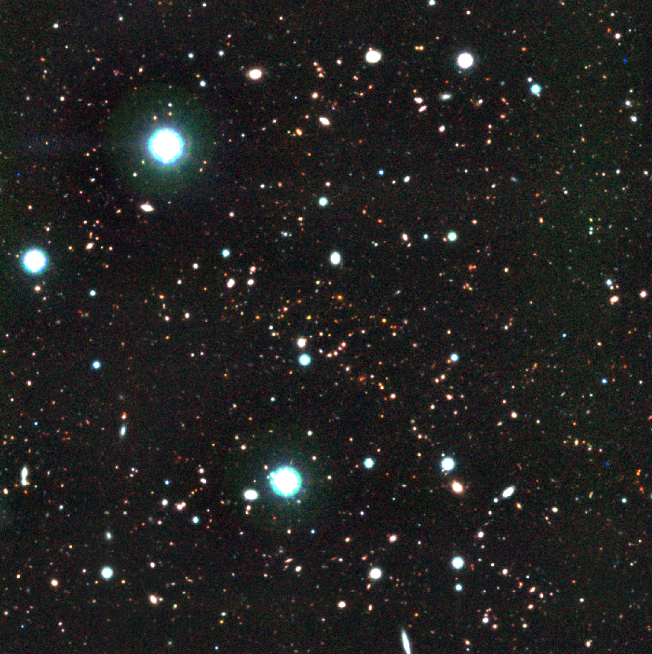}
\caption{A $6 \times 6$~arcmin$^{2}$ 3-colour image (using YJKs data) from the VIDEO survey (Jarvis et al. in prep). (Best viewed in colour.)}\label{fig:video_image}
\end{figure*}

The VIDEO survey will reach the following $5\sigma$ (2~arcsec
point source) depths; $Z=25.7$, $Y=24.6$, $J=24.5$, $H=24.0$ and
$K_{\rm s}=23.5$ (all AB magnitudes), and will use just over
200~nights of observing time over the next five years. In terms of the
galaxies which will be detected in the VIDEO survey, these depths
correspond to an $L^{\star}$ elliptical galaxy up to $z\sim4$ and
$0.1~L^{\star}$ galaxy up to $z\sim1$, thus it sits naturally between
the Ultra-VISTA and VIKING surveys.The SERVS depths are slightly
deeper in real terms with the proposed depth being able to detect a
$L^{\star}$ elliptical galaxy up to $z\sim 5$. A three-colour image over a $6\times 6$~arcmin$^2$ portion of one of the VIDEO fields is shown in Fig.~\ref{fig:video_image}.

This combination of depth and area ensures that with the VIDEO survey
and SERVS we will be able to trace galaxy formation and evolution over the majority of cosmic history and over all environmental densities. Some of the main science goals are; 

\begin{itemize}

\item to trace the evolution of galaxies from the earliest epochs until the present day. The depth and breadth of the VIDEO survey will enable galaxy properties to be measured as a function of environmental density,  allowing the detailed study of how the the environmental richness may affect the properties of the galaxies. Furthermore, the combination of VIDEO data with the wealth of multi-wavelength data over these regions of sky, will allow the star-formation rate and AGN activity to be be measured and linked to the stellar mass in the galaxies.

\item to measure the clustering and space density of the most massive galaxies at $z>5$ where we expect to detect $\sim 300$ galaxies with $M>10^{11}$~M$_{\odot}$ with $\sim 100$ of these at $z>6$.

\item to trace the evolution of galaxy clusters from the formation epoch until the present day\cite{vB06}. The depth of VIDEO ensures that we will be able to trace the bright end of the cluster luminosity function to $z\sim 3$, while its area should provide a sample of $\sim 70$ clusters with $M>10^{14}$M$_{\odot}$ at $z>1$, with around 15 of these expected to be at $z>1.5$.

\item to quantify the accretion activity over the history of the Universe. The depth and area, along with the filter combination will allow the detection of the highest redshift quasars, and this place tight constraints on the quasar luminosity function at $z>6.5$, when combined with  wider and shallower surveys, such as VIKING. Furthermore, VIDEO data can be combined with {\sl Spitzer} and {\sl Herschel} data to place constraints on both the obscured AGN and star-forming galaxies, all the way out to $z\sim6$ if AGN host galaxies have typical luminosities of $2-3~L^{\star}$\cite{Jarvis01,Herbert11}

\item to help reduce the systematic uncertainties in using Type Ia supernovae for cosmology. The near-infrared imaging data provided by the VIDEO survey will enable much more accurate measurements of the properties of SNe hosts, thus allowing dust extinction to be measured, along with the galaxy type and quantification of the stellar populations.

\end{itemize}

Over the coming years the VIDEO fields will be targeted by a range of other major astronomical facilities. At optical wavelengths the VLT Survey Telescope (VST), the Dark Energy Survey (DES) and SkyMapper all have plans to observe these fields to varying depths and a variety of optical filters. This will ensure that the near-infrared VIDEO data can be combined with some of the best optical imaging data available.

The VIDEO survey will therefore provide the ideal data to be combined with any deep extragalactic observations with the SKA-pathfinder telescopes in both continuum and H{\sc i}. Crucially, VIDEO will have the depth to detect low-mass galaxies which would dominate the $\ltsim1$ SKA H{\sc i} surveys, thus providing photometric redshifts and important information on the stellar populations in these galaxies.

\section{Herschel \& SCUBA-2}

To obtain a full picture of the evolutionary history of the Universe one needs to combine data from all wavelengths that are emitted by a range of processes. The power of {\sl Herschel} and SCUBA-2 will provide the crucial measurement of the obscured activity in the Universe.

I describe some of the science results from one of the large {\sl Herschel} surveys, namely {\sl Herschel}-ATLAS\cite{Eales10} in the next section. However, it is also worth mentioning that the {\sl Herschel} Multi-tiered Extragalactic Survey (HerMES\citep{Oliver10}) Guaranteed Time survey is surveying the VIDEO fields with the aim of addressing similar science but to higher redshifts and greater depth, allowing a view of the obscured activity at $z>1$ to be formed. A joint call for papers wanting to combine HerMES, VIDEO and SERVs data has just been announced at the time of writing.

The SCUBA-2 Cosmology Legacy Survey (S2CLS) will also survey the VIDEO/SERVs fields. It is worth noting that the VIDEO and SERVs depths are such that they will be able to detect many of the SCUBA sources and future radio facilities such as LOFAR, ASKAP and MeerKAT should detect all of them.

In the following section I highlight a selection of the science already achieved by the {\sl Herschel}-ATLAS survey to give a broad aspect of what can be achieved, not only by deep narrow surveys, but also wider medium-deep surveys.

\subsection{{\sl Herschel}-ATLAS}\label{sec:HATLAS}
{\sl Herschel}-ATLAS (H-ATLAS) has the largest time allocation ($\sim 600$~hours) of any programme from the initial call for Open Time Key Projects on {\sl Herschel}, it will cover $\sim 550$~square degrees in five far-infrared and submm wavebands from 100-550$\mu$m.  H-ATLAS has a diverse range of science goals, including:

\begin{itemize}

\item The H-ATLAS will be the first submm survey large enough to detect a
significant number of galaxies in the nearby universe, enabling the study of the evolution of obscured star formation \citep{Dye10} and dust \citep{Dunne11}, how star formation may depend on the environmental density \citep{Dariush11} and in what mass of halo do submillimetre galaxies reside \citep{Guo11}.
By carrying out the
survey in the fields surveyed in the SDSS, 2dFGRS and the GAMA
survey \citep{Driver09,Driver11}, a significant fraction ($\sim 20$\%) already have spectroscopic redshifts and closer to 40\% with either photometric or spectroscopic redshifts \citep[see e.g.][]{Smith11a}.

\item H-ATLAS is being used to investigate the relationship
between the star formation and the black hole activity and how this
relationship changes over time. By observing a very large sample of
quasars, Bonfield et al.\citep{Bonfield11} show that the far-infrared luminosity and accretion luminosity are correlated with little scatter, suggesting a direct physical connection between star formation and accretion activity. The volume covered by H-ATLAS also allows the study of other rare objects, such as radio galaxies, where Hardcastle et al.\citep{Hardcastle10} have shown that, at least for the low-luminosity FRI-type radio galaxies, there is no evidence for an increased level of star formation in the host galaxy, in agreement with optical studies\cite{Herbert10}.

\item The strong negative $k-$correction at submm wavelengths means that large area shallow surveys such as H-ATLAS provide an excellent way of finding large numbers of strong gravitational lenses. This has been dramatically shown  with the early result from both H-ATLAS \cite{Negrello10} and HerMES\cite{Conley11}. The magnification of the far-infrared emission from these lenses is extremely useful as it allows their molecular composition to be studied in detail with current instrumentation \cite{Frayer11, Lupu11, Omont11, Cox11}.

\item The large area, coupled with the large redshift range that far-infrared and submillimetre surveys extend over, also ensures that H-ATLAS (and HerMES) are ideally suited to tracing large-scale structure in the Universe and linking the obscured star-forming galaxies to the underlying dark matter haloes\cite{Maddox10,Amblard11}
\end{itemize}

Therefore, the combination of the multi-wavelength data from the $\sim 100$~square degree fields such as H-ATLAS, VIKING and KIDS with the deeper, smaller fields such as HerMES, VIDEO and SERVS will provide a powerful data set for understanding the evolution of galaxies across cosmic time and as a function of environmental density.

\section{The role of MeerKAT}

The Meer Karoo Array Telescope (MeerKAT\cite{Jonas09}) is the South African precursor to the Square Kilometre Array. It will have 64 13.5~m offset-Gregorian dishes operating with wide-band receivers, eventually with three frequency bands; $580 \ltsim \nu \ltsim 1000$~MHz, $1000 \ltsim \nu \ltsim 1750$~MHz and $8 \ltsim \nu \ltsim 14.5$~GHz, with a field of view of $\sim 0.8$~degree$^{2}$ at 1400~MHz. 
Although the field-of-view is not competitive with the Australian SKA Pathfinder (ASKAP\cite{Schinckel09}), MeerKAT's sensitivity far outstrips ASKAP, because of this MeerKAT is ideally suited to relatively narrow surveys with large depth, making it the ideal complementary facility to the deep and narrow surveys discussed above, namely VIDEO, SERVS and HerMES.
In the following sections I discuss the approved MeerKAT radio continuum survey, MIGHTEE, and also the synergies with two other surveys, namely the deep H{\sc i} survey LADUMA and the deep high-frequency CO survey at $z>6$, MESMER.

\subsection{The MIGHTEE survey}

\begin{figure*}[ht!]
 \centering
 \includegraphics[width=0.9\linewidth]{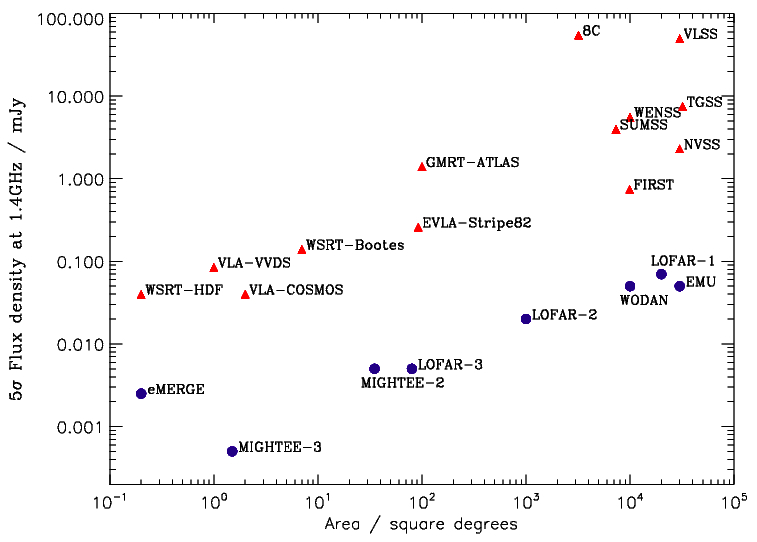}
 \caption{Area versus $5\sigma$ flux-density for a range of existing (red triangles) and future (blue circles) radio surveys. Those survey conducted at frequencies other than 1.4~GHz, a spectral index of $\alpha= 0.8$ ($S\propto \nu^{-\alpha}$) has been used to estimate the the corresponding 1.4~GHz flux-density. 8C\cite{Rees90}, VLSS\cite{Cohen07}, TGSS\cite{TGSS}, NVSS\cite{NVSS}, FIRST\cite{FIRST}, SUMSS\cite{SUMSS}, WENSS\cite{WENSS}, EVLA-Stripe 82\cite{Hodge2011}, GMRT-ATLAS\cite{GMRT-ATLAS}, WSRT-Bootes\cite{WSRT-Bootes}, VLA-COSMOS\cite{VLA-COSMOS}, VLS-VVDS\cite{VLA-VVDS}, WSRT-HDF\cite{WSRT-HDF}, eMERGE\cite{eMERGE}, LOFAR Tiers 1, 2 and 3\cite{LOFAR}, EMU\cite{Norris11} and MIGHTEE Tiers 2 and 3 (this article). }\label{fig:areaflux}
\end{figure*}

The MeerKAT International Giga-Hertz Tiered Extragalactic Exploration (MIGHTEE) survey will utilize MeerKAT's unique capabilities, namely high-resolution coupled with a large survey speed, to undertake a tiered survey strategy at 1.4 GHz. The combination of resolution and sensitivity (outperforming both ASKAP and EVLA; see Figure~\ref{fig:areaflux}) will allow us to make the most precise measurement of the radio luminosity function for radio-loud and radio-quiet AGN and star-forming galaxies over the full range in radio luminosity. 

The original proposal had wide tier spanning 1000 square degrees down to an rms flux-density limit of $S_{\rm 1.4GHz} = 5\,\mu$Jy, which was not approved by the time allocation committee. This would have provided the combination of depth and area to pin down the luminosity function at bright radio luminosities out to $z \sim 5$, but can also be used to find rare AGN in the epoch of reionization, undertake detailed studies of galaxy clusters out to $z \sim 1$, provide the data to investigate associated H{\sc i} absorption in and around radio galaxies, radio quiet quasars and starburst galaxies, and produce Faraday rotation maps. The ASKAP-EMU survey\cite{Norris11} will contribute greatly to these questions, but the MIGHTEE Tier-1 would have filled an important part of the parameter space from the all-sky EMU survey to the MIGHTEE Tiers-2 and 3.

The approved Tier-2 covers 35 square degrees to an rms flux-density of $1\,\mu$Jy; it will allow detailed studies of the distant Universe and the evolution of the lower-luminosity radio source populations into the epoch of reionization. 
The chosen fields will ensure that the maximum scientific return is obtained by combining the deep radio continuum data with the best multi-wavelength data sets from X-ray through to far-infrared wavelengths, namely the whole area covered by the lowest tier of the HerMES survey which include the SWIRE data, the VIDEO data and the SERVS data.

Tier-3 will utilize the high-resolution and increased bandwidth to push the telescope to its limit (0.1$\,\mu$Jy in a single pointing). We will also be able to carry out studies of the evolution of Milky Way type galaxies up to $z \sim 4$  and test techniques to increase dynamic range in preparation for the full SKA. This tier will be carried out commensally with the Looking at the Distant Universe with the MeerKAT Array (LADUMA) 21-cm H{\sc i} survey.

Tiers 2 and 3, together with EMU, will enable the star-formation and accretion activity to be traced across 90\% of cosmic time, at radio wavelengths that are not attenuated by dust extinction. However, the radio continuum observations do not provide any redshift information, therefore this science can only be achieved with the redshift information gleaned from the optical and near-infrared imaging from DES, VST, VIDEO and SERVS using photometric redshifts, or spectroscopic redshifts through a large coordinated campaign (see section on SALT below).

Work done thus far in this area is gradually gaining momentum as radio continuum surveys reach to deeper depths allowing the far-infrared--radio correlation\cite{Ivison10,Jarvis10} and therefore the  star formation\cite{Dunne09,Ibar09} to be traced, along with the low-luminosity AGN activity\cite{Smolcic09,McAlpine11}.

\subsection{Synergies with LADUMA}

One of the key aims of the SKA and its precursors is to measure the evolution of neutral hydrogen through observations of the 21~cm emission (and absorption). The LADUMA survey\cite{Holwerda10} will target a single patch of sky within the extended Chandra Deep Field South (ECDFS), which is again a HerMES, VIDEO and SERVs field. This will allow the gas phase of the galaxies to be linked to the stellar populations using optical and near-infrared imaging data, the dust composition using the HerMES data. The radio continuum emission will be carried out commensally. Furthermore, optical imaging and follow-up optical spectroscopic redshifts with SALT (see below), will allow the stacking of the radio spectral line data to push below the rms noise of the radio data. Allowing the neutral hydrogen content to be measured, in a statistical sense, down to much lower H{\sc i} masses\cite{Lah09}.

\subsection{Synergies with MESMER and the Epoch of Reionization}

The MeerKAT Search for Molecules in the Epoch of Reionization (MESMER\cite{Heywood11}) is another MeerKAT survey that will be carried out over the Extended Chandra Deep Field South (along with other targeted fields) at the high-frequency band of MeerKAT. The key drivers for this survey will be to detect and study the molecular content of galaxies within the epoch of reionization with CO. The survey will cover the redshift range $6.95<z<9.98$ with a predicted source density of $\sim 180$ galaxies per square degree\cite{Heywood11}. Obviously such deep data at these high frequencies can be combined with the lower frequency data from MIGHTEE and LADUMA to obtain additional information on the source structure, utilising the factor of 10 higher resolution at 14~GHz compared to 1.4~GHz and spectral index information. 

\section{The role of the Southern African Large Telescope (SALT)}\label{sec:SALT}

All of the surveys discussed above are either solely in the southern hemisphere or have a large component accessible to southern-hemisphere based observatories. Thus, the recent successful re-commissioning of SALT opens up a great opportunity for high-impact science to be led from Africa. I highlight below some key contributions that SALT could make to the surveys discussed previously.

The multi-object spectrograph currently offered on SALT makes it an ideal instrument for the follow-up of faint targets with high surface density. Therefore, an obvious follow-up programme would be to obtain spectroscopic redshifts for the galaxy populations present in the relatively narrow-field VIDEO and Ultra-VISTA surveys. Spectroscopic surveys have already been carried out over sub-sections of these fields, predominantly using the VIMOS instrument on the VLT\citep{zCOSMOS, VVDS}. However, SALT offers a higher sensitivity at the far-blue end of the optical spectrum, allowing unique science to be achieved.

Of particular interest, and a topic that is closely linked to the MeerKAT science goals, is to gain spectroscopic redshifts and emission line properties of the radio sources in the VIDEO fields, utilizing the already available radio continuum data\cite{VLA-VVDS}. This would allow an in-depth study of the evolution of the low-excitation and high-excitation radio sources within this field. This would result in a much clearer picture of the evolution in the accretion mode of supermassive black holes in galaxies over cosmic time. Linked to this, the extremely deep near-infrared data from SERVS and VIDEO will also provide the possibility of efficient selection of high-redshift radio sources with the aim of probing the epoch of reionization\cite{Jarvis09}, another key aim of the SKA. Moreover, a general spectroscopic redshift  survey of galaxies over field which include data from {\sl Herschel}  would allow, not only the properties of far-infrared sources to be studied in detail, but also the stacking of far-infrared data to probe below the classical confusion limit of the {\sl Herschel} surveys\cite{Bonfield11}.

One of the key aims of LADUMA is to probe to low H{\sc i} masses through stacking at the positions of galaxies of known redshifts. 
Well before MeerKAT begins operation, slits on the multi-object spectrograph could be placed on galaxies with photometric redshifts covering the range in redshift of the LADUMA H{\sc i} survey.  Such data would provide the necessary information to allow an increasingly efficient selection of targets for follow-up spectroscopy in preparation for LADUMA by investigating the combination of colours which allow the selection of emission-line objects, for which H{\sc i} would presumably be more abundant.

Such projects would place SALT at the forefront of the follow-up of large international surveys and may provide the African community with large observational data sets that it generally would not normally have immediate access to.

\section{Acknowledgments}

First, I would like to thank the MEARIM-II organisers for inviting me to an excellent meeting which really showed the breadth and depth of science being conducted in the Middle East and Africa.
I'd like to thank all of the people who have and continue to contribute to the surveys that I have discussed here, particularly Dave Bonfield who has worked tirelessly in producing the data products for VIDEO, my co-PI of the MIGHTEE survey, Kurt van der Heyden, the PI of SERVS Mark Lacy and all of the people involved in the {\sl Herschel} surveys.

\bibliographystyle{plainnat}
\bibliography{references}

\end{document}